# *MatBase* Algorithm for Translating Entity-Relationship Data Models into (Elementary) Mathematical Data Model Schemes

**Christian Mancas* and Diana Christina Mancas**

*Mathematics and Computer Science Dept., Ovidius University at Constanta, Romania*

***Corresponding Author:** Christian Mancas, Mathematics and Computer Science Department, Ovidius University at Constanta, Bd. Pipera 1/U, Voluntari, IF, Romania.

## Abstract

This paper presents a pseudocode algorithm for translating Entity-Relationship data models into (Elementary) Mathematical Data Model schemes. We prove that this algorithm is linear, sound, complete, and optimal. As an example, we apply this algorithm to an Entity-Relationship data model for a teaching sub-universe. We also provide the main additional features added to the implementation of this algorithm in *MatBase*, our intelligent knowledge and database management system prototype based on both the Entity-Relationship, (Elementary) Mathematical, and Relational Data Models.

## Abbreviations

DBMS = Database Management System.
db(s) = database(s).
(E)MDM = (Elementary) Mathematical Data Model.
E-R = Entity-Relationship.
ERD(s) = Entity-Relationship diagram(s).
ERDM = Entity-Relationship Data Model.
ERDS = Entity-Relationship diagram set.
RDBMS(s) = Relational Database Management System(s).
RDM = Relational Data Model.

## Introduction

The Entity-Relationship (E-R) Data Model (ERDM) [1-3] has proved for decades to be the best initial database (db) conceptual design tool, as its graphic E-R Diagrams (ERDs) are easy to understand by customers as well. In [3], we defined E-R data models as triples <*ERDS, ARS, ICSD*>, where *ERDS* is a set of *ERDs*, *ARS* is an Associated Restriction Set, and *ICSD* is an Informal Corresponding Sub-universe Description. The associated restrictions, which correspond to the business rules governing the modeled sub-universes, are of the following five types: inclusions between object sets (e.g., $TEACHERS \subseteq EMPLOYEES$), ranges of the attributes (e.g., *Weekday* between 1 and 7), compulsory (not null) attribute values (e.g., *Name* compulsory), minimal uniqueness of attributes (e.g., *SSN* unique) and attribute concatenations (e.g., $Room \cdot Weekday \cdot StartHour$ unique), and other restriction types (e.g., no student may be simultaneously present in two classrooms).

Once agreed with customers, E-R data models may be directly translated into Relational Data Model (RDM) schemes [3-5] and implemented with a Relational Database Management System (RDBMS, e.g., Oracle Database, MS SQL Server, IBM DB2, etc.), with restrictions translated into db constraints. As RDBMSs provide only six (relational) constraint types (i.e., domain/range, not null, default value, unique key, foreign key, and tuple/check), all other (nonrelational) constraints must be enforced by the software applications managing the corresponding dbs.

However, E-R data models are not formal, so prone to errors and omissions. Consequently, it is preferable to first translate them into a formal, higher-level conceptual data model, for both validation and refinement, and only then translate the corresponding schemes into relational ones and associated nonrelational constraint sets. One such model is our (Elementary) Mathematical Data one ((E)MDM) [6, 7].

*MatBase* [8, 9] is our intelligent knowledge and DBMS prototype, based on both (E)MDM, RDM, and ERDM, which currently has two versions - one, for small and medium dbs, developed in MS Access, and one, for large dbs, in MS SQL Server and C#.NET.

The next Section introduces and characterizes the pseudocode algorithm used by *MatBase* to translate E-R data models into (E)MDM schemes. The third one presents and discusses the result of applying this algorithm to an E-R data model from [3]. The paper ends with conclusion and a reference list.

## Materials and Methods

Figures 1 to 3 show the *MatBase* algorithm *A*1 for translating E-R data models into (E)MDM schemes.

***Proposition*** (Algorithm *A*1's characterization)

***Algorithm A1 is:***

(i) linear, having complexity $O(S + A + C)$, where *S* is the total number of *ERDS* object sets, *A* is the total number of their attributes, including the roles of the relationship-type sets, and *C* is the total number of associated restrictions;
(ii) sound;
(iii) complete;
(iv) optimal.

***Proof:***

(i) Obviously, *A*1 performs $E + CS + R + RA$ steps in its first outer loop from Fig. 1 (where *E* is the total number of entity-type sets, *CS* the one of computed sets, *R* the one of relationship-type sets from *ERDS*, and *RA* is the total number of relationship roles) plus *NR* steps in its final outer loop (*NR* being the total number of associated nonrelational constraints); the loop in the method *addSet* from Fig. 2 is performed *IC* times (*IC* being the total number of inclusions between *ERDS* object sets); the first loop in the method *completeScheme* from Fig. 3 is performed *SF* times (*SF* being the total number of *ERDS* structural functions, i.e., functional relationships, be they fundamental or computed); the second one is performed $EN + U$ times (where *EN* is the total *ERDS* ellipses number, be they fundamental or computed, and *U* the total number of corresponding unique attributes); the third one is performed *CR* times (*CR* being the total number of compulsory, i.e., not null, restrictions); the fourth one is performed *CU* times (*CU* being the total number of concatenated minimal uniqueness restrictions); finally, the fifth one is performed *TR* times (*TR* being the total number of tuple, i.e., check, restrictions); consequently, as, by the E-R data model definition, $S = E + R$

+ *CS*, *A* = *RA* + *SF* + *EN* and *C* = *NR* + *IC* + *CR* + *U* + *CU* + *TR*, it follows that *A*1 always performs exactly *S* + *A* + *C* steps, so it never infinitely loops.

(ii) As *A*1 always outputs only sets, mappings, and constraints, it is sound.

(iii) As *A*1 always translates any *ERDS* entity, computed, or relationship type set into a(n) (E)MDM set, any relationship role, structural function, or attribute, be they fundamental or computed, into a mapping, and any associated restriction into a constraint, it is also complete.

(iv) As *A*1 translates any object set, attribute, and associated restriction only once, in the minimum possible number of steps, it is also optimal. *Q.E.D.*

**Algorithm *A*1 (*E-R data models translation into (E)MDM schemas*)**

*Input*: an E-R data model *M*; *Output*: the corresponding (E)MDM db scheme *S*; *Strategy*:

*loop for all* E-RDs *D* from *M*<br>
  *loop for all* rectangles *R* in *D* (in bottom-up order, from only referenced, non-referencing object sets to non-referenced, only referencing ones)<br>
    *if R* is not a computed one *then addSet*; *else* add to *S* the corresponding *R*'s definition;<br>
    *end if;*<br>
  *end loop*;<br>
  *loop for all* diamonds *R* in *D* (in the same as above bottom-up order)<br>
    *addSet*;<br>
    *loop for all R*'s roles *r*<br>
      add to *R*'s scheme a role (canonical Cartesian projection) $r \to U$, where *U* is the set corresponding to the rectangle or diamond with which *r* connects *R*;<br>
    *end loop*;<br>
  *end loop*;<br>
*end loop*;<br>
*loop for all* associated non-relational restrictions *nrr*<br>
  formalize *nrr* as a constraint and add it to *S*;<br>
*end loop*;<br>
*End Algorithm A*1;

***Figure 1:*** Algorithm *A*1 (Translate E-R data models into (E)MDM schemes).

***Method addSet***

*if* there is no set *R then* add set *R*;<br>
  even if *R* has no surrogate key, add object identifier $x \leftrightarrow$ NAT(*n*), where *n* is the power of 10 in *R*'s maximum cardinality restriction;<br>
  *loop for all* sets *T* with $R \subseteq T$<br>
    add constraint $R \subseteq T$;<br>
  *end loop*;<br>
  *completeScheme*;<br>
*endif;*<br>
*End Method addSet;*

***Figure 2:*** Method *addSet* of Algorithm *A*1 from Fig. 1.

## Results and Discussion

For example, let us consider the E-R data model from [3], consisting of the E-RD from Figure 2.10 (page 45, see Fig. 4) and the associated restriction set made of the range ones R01 to R19 (page 53), compulsory ones R20 to R27 (page 56), uniqueness ones R28 to R36 (page 58), and other type ones R37 to R41 (page 59, see Fig. 5).

**Method completeScheme**

*loop for all* arrows $A$ from $R$ to $U$, except for set inclusion-type ones
  *if* $A$ is not computed *then* add a structural function $A : R \rightarrow U$;
  *else* add the definition of the computed function $A$; *end if;*
*end loop*;
*loop for all* ellipses $e$ connected to $R$, except for the surrogate key one
  *if* $e$ is not a computed one *then* add to $R$'s scheme attribute $e \rightarrow V$ (or $e \leftrightarrow V$, if $e$ has an associated uniqueness restriction), where $V$ is the value set corresponding to $e$'s range restriction;
  *else* add to $R$'s scheme the definition of the computed attribute $e$;
  *end if;*
*end loop*;
*loop for all* compulsory restrictions $c$ associated to $R$
  add *to R's* scheme corresponding total constraints;
*end loop*;
*loop for all* concatenated uniqueness restrictions $u$ associated with $R$
  add *to R's* scheme corresponding *u key* constraint;
*end loop*;
*loop for all* tuple-type restrictions $t$ associated to $R$
  add *to R's* scheme corresponding $(\forall x \in R)t(x)$ constraint;
*end loop*;
*End Method completeScheme*;

***Figure 3:*** Method *completeScheme* of Algorithm *A*1 from Fig. 1.

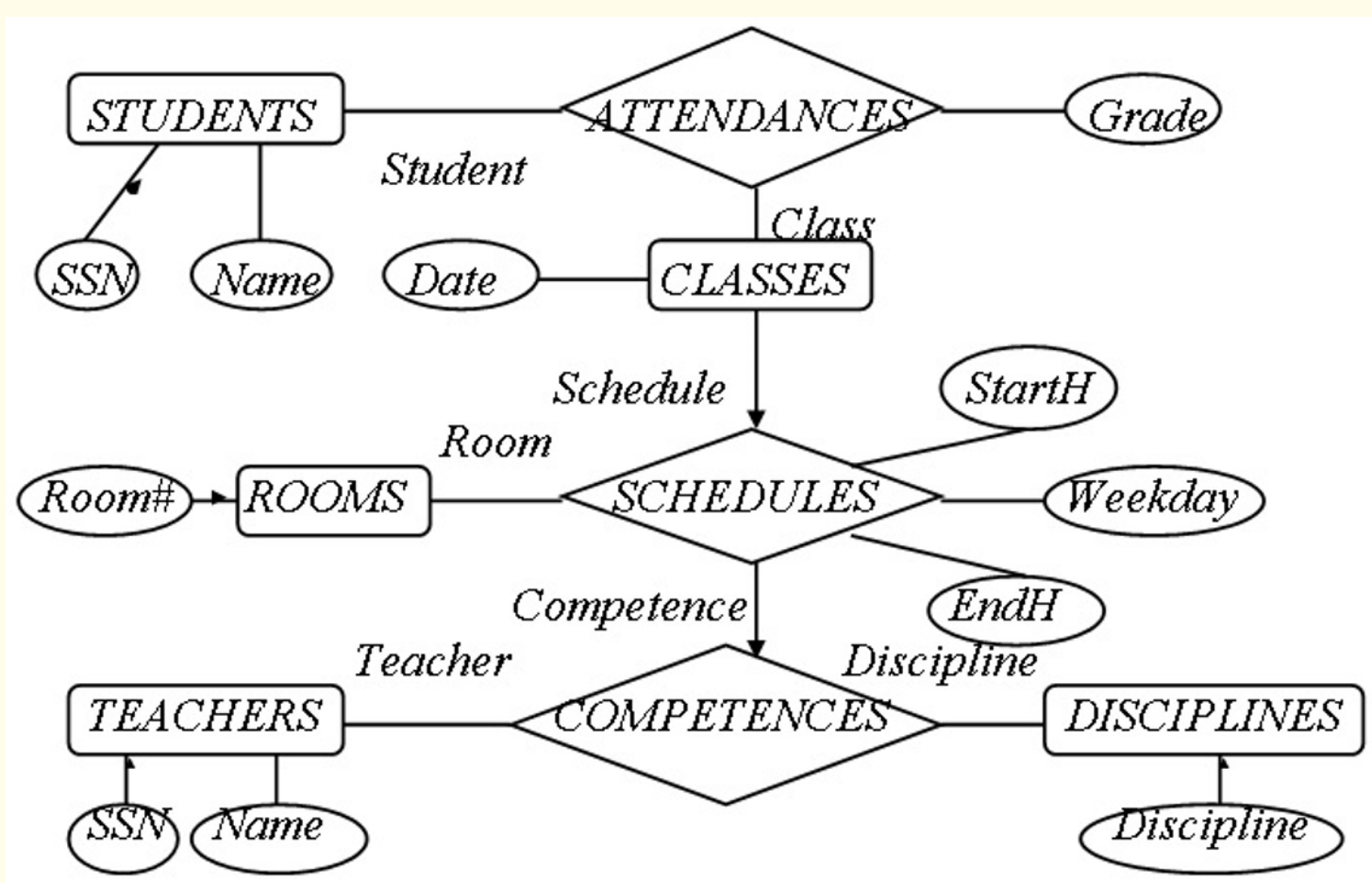

***Figure 4:*** The ERD of a teaching sub-universe.

| | |
|---|---|
| ***STUDENTS*** | |
| $max(card(STUDENTS)) = 10^5$ | (R01) |
| *SSN*: [1000101000000, 8991231999999] | (R02) |
| *Name*: ASCII(255) | (R03) |
| *Compulsory*: *SSN*, *Name* | (R20) |
| *Uniqueness*: *SSN* | (R28) |
| ***TEACHERS*** | |
| $max(card(TEACHERS)) = 10^3$ | (R04) |
| *SSN*: [1000101000000, 8991231999999] | (R05) |
| *Name*: ASCII(255) | (R06) |
| *Compulsory*: *SSN*, *Name* | (R21) |
| *Uniqueness*: *SSN* | (R29) |
| ***DISCIPLINES*** | |
| $max(card(DISCIPLINES)) = 10^3$ | (R07) |
| *Discipline*: ASCII(128) | (R08) |
| *Compulsory*: *Discipline* | (R22) |
| *Uniqueness*: *Discipline* | (R30) |
| ***ROOMS*** | |
| $max(card(ROOMS)) = 10^3$ | (R09) |
| *Room#*: $[1, 10^4]$ | (R10) |
| *Compulsory*: *Room#* | (R23) |
| *Uniqueness*: *Room#* | (R31) |
| ***CLASSES*** | |
| $max(card(CLASSES)) = 10^4$ | (R11) |
| *Date*: [01/10/2010, *SysDate*()] | (R12) |
| *Compulsory*: *Date*, *Schedule* | (R24) |
| *Uniqueness*: *Date* • *Schedule* | (R32) |

***Figure 5:*** The restriction set associated with the ERD from Fig. 4.

Applying Algorithm *A*1 to this E-R data model results in the (E)MDM scheme shown in Fig. 6.

Please note that, according to the (E)MDM implicit conventions [7], uniqueness restrictions R35 and R36 are not explicitly figured, whereas a new such constraint (i.e., R42: *Room* • *Competence*) is added to the relationship *SCHEDULES*. Algorithm A2 from [3] ("Assisting validation of the initial E-R data model") should discover that this constraint does not correspond to any existing business rule governing this sub-universe (as a teacher may teach a same discipline in a same classroom several times, provided it is done at different hour intervals or/and weekdays) and hence discover that, in fact, *SCHEDULES* is not a relationship-type object set but an entity-type one and, hence, discards R42.

Translations of everything except the nonrelational constraints are automatically performed by *MatBase*. For the nonrelational constraints, *MatBase* asks its users to provide the corresponding formalized constraints one by one. A truly useful and powerful artificial intelligence (AI) tool would be one for converting plain English into first-order predicate logic expressions, whenever possible!

In fact, actual *MatBase* algorithms *A*1 are more complex. For example:

(i) if any cardinality restriction is missing, then *MatBase* assumes the maximum one of the corresponding DBMS;
(ii) if a cardinality exceeds the maximum available, it replaces it with that maximum;
(iii) if a fundamental ellipse lacks its range, it assumes ASCII(255) for it;

| | |
|---|---|
| ***SCHEDULES*** | |
| $max(card(SCHEDULES)) = 10^5$ | (R13) |
| $Weekday: [1, 5]$ | (R14) |
| $StartH: [8, 19]$ | (R15) |
| $EndH: [9, 20]$ | (R16) |
| $Compulsory: Weekday, StartH, EndH, Room, Competence$ | (R25) |
| $Uniqueness: Room \bullet Weekday \bullet StartH$ | (R33) |
| $Room \bullet Weekday \bullet EndH$ | (R34) |
| $StartH < EndH$ | (R37) |
| ***ATTENDANCES*** | |
| $max(card(ATTENDANCES)) = 10^9$ | (R17) |
| $Grade: [1, 10]$ | (R18) |
| $Compulsory: Student, Class$ | (R26) |
| $Uniqueness: Student \bullet Class$ | (R35) |
| ***COMPETENCES*** | |
| $max(card(COMPETENCES)) = 10^4$ | (R19) |
| $Compulsory: Teacher, Discipline$ | (R27) |
| $Uniqueness: Teacher \bullet Discipline$ | (R36) |
| No teacher may be simultaneously present in more than one room. | (R38) |
| No student may be simultaneously present in more than one room. | (R39) |
| No room may simultaneously host more than one class. | (R40) |
| There may not be two people (be they teachers or students) having same SSN. | (R41) |

***Figure 5:*** (Continued).

(iv) if a computed set or mapping lacks its computing expression, it asks users for it and if it does not get one it ignores it;

(v) *MatBase* automatically adds totality constraints to any role and object identifier;

(vi) if a fundamental (i.e., not computed) object set has no compulsory mapping defined on it, then *MatBase* adds a total one called *Compulsory*, taking values from ASCII(255);

(vii) if a relationship-type set has no structural key (i.e., a key made up of only its roles), then *MatBase* adds a key made of all its roles;

(viii) if a binary relationship-type set $R = (f \to S, g \to T)$ has $f$ unique, then *MatBase* replaces it by the structural function $R : S \to T$; if $g$ is unique, then it replaces it by $R : T \to S$; if both $f$ and $g$ are declared as unique, then it replaces it by $R : S \leftrightarrow T$ or $R : T \leftrightarrow S$, according to the corresponding users' choice;

(ix) if a fundamental object set has no uniqueness restriction, then *MatBase* adds a one-to-one and total mapping called *Unique-Mapping* defined on it and taking values from ASCII(255).

As expected, in any of the above situations, *MatBase* displays corresponding error, warning, and/or information messages.

***STUDENTS***
$x \leftrightarrow$ NAT(5), total
*SSN* $\leftrightarrow$ [1000101000000, 8991231999999], total
*Name* $\rightarrow$ ASCII(255), total
***TEACHERS***
$x \leftrightarrow$ NAT(3), total
*SSN* $\leftrightarrow$ [1000101000000, 8991231999999], total
*Name* $\rightarrow$ ASCII(255), total
***DISCIPLINES***
$x \leftrightarrow$ NAT(3), total
*Discipline* $\leftrightarrow$ ASCII(128), total
***ROOMS***
$x \leftrightarrow$ NAT(3), total
*Room#* $\rightarrow$ [1, $10^4$], total
***CLASSES***
$x \leftrightarrow$ NAT(4), total
*Date* $\rightarrow$ [01/10/2010, *SysDate*()], total
*Schedule* : *CLASSES* $\rightarrow$ *SCHEDULES*, total
R32: *Date* • *Schedule* key
***COMPETENCES*** = (*Teacher* $\rightarrow$ *TEACHERS*, *Discipline* $\rightarrow$ *DISCIPLINES*)
$x \leftrightarrow$ NAT(4), total

***SCHEDULES*** = (*Room* $\rightarrow$ *ROOMS*, *Competence* $\rightarrow$ *COMPETENCES*)
$x \leftrightarrow$ NAT(5), total
*Weekday* $\rightarrow$ [1, 5], total
*StartH* $\rightarrow$ [8, 19], total
*EndH* $\rightarrow$ [9, 20], total
R33: *Room* • *Weekday* • *StartH* key
R34: *Room* • *Weekday* • *EndH* key
R37: $(\forall x \in SCHEDULES)(StartH(x) < EndH(x))$
***ATTENDANCES*** = (*Student* $\rightarrow$ *STUDENTS*, *Class* $\rightarrow$ *CLASSES*)
$x \leftrightarrow$ NAT(9), total
*Grade* $\rightarrow$ [1, 10]

R38: $(\forall x,y \in SCHEDULES)(Teacher(Competence(x)) = Teacher(Competence(y)) \wedge Weekday(x) = Weekday(y) \wedge Room(x) = Room(y) \Rightarrow StartH(x) \neq StartH(y))$
R39: $(\forall u,v \in ATTENDANCES)(\forall x,y \in SCHEDULES)(Student(u) = Student(v) \wedge x = Schedule(Class(u)) \wedge y = Schedule(Class(v)) \wedge Weekday(x) = Weekday(y) \wedge Room(x) = Room(y) \Rightarrow StartH(x) \neq StartH(y))$
R40: $(\forall u,v \in CLASSES)(\forall x,y \in SCHEDULES)(Schedule(u) = Schedule(v) \wedge x = Schedule(u) \wedge y = Schedule(v) \wedge Weekday(x) = Weekday(y) \wedge Room(x) = Room(y) \Rightarrow StartH(x) \neq StartH(y))$
R41: $(\forall x \in STUDENTS)(\forall y \in TEACHERS)(SSN(x) \neq SSN(y))$

***Figure 6:*** The (E)MDM scheme obtained from the E-R data model from Fig. 4 and 5.

## Conclusion

We presented a linear, sound, complete, and optimal pseudocode algorithm for translating E-R data models into (E)MDM schemes used by both versions of our intelligent DBMS prototype *MatBase*. Obviously, this algorithm may be also manually used by db and/or software architects.

We provided an example of applying it to a teaching sub-universe.

We also described the powerful additional features of its actual implementations that are aimed at obtaining the highest possible quality of data modeling.

## Conflict of interest

The authors declare that the research was conducted in the absence of any commercial or financial relationships that could be construed as a potential conflict of interest.

## Acknowledgements

This research was not sponsored by anybody and nobody other than its authors contributed to it.